\documentclass{aastex}

\usepackage{amsmath}
\usepackage{amstext}
\usepackage{graphicx}
\usepackage[caption=false]{subfig}

\begin{document}

\title{A NEW MODEL FOR MIXING BY DOUBLE-DIFFUSIVE CONVECTION (SEMI-CONVECTION). III. \\ THERMAL AND COMPOSITIONAL TRANSPORT THROUGH NON-LAYERED ODDC}
\author{Ryan Moll and Pascale Garaud}
\affil{Department of Applied Mathematics and Statistics, Baskin School of Engineering, University of California Santa Cruz, CA 95064, USA}
\and
\author{Stephan Stellmach}
\affil{Institut f{\"u}r Geophysik, Westf{\"a}lische Wilhelms-Universit{\"a}t M{\"u}nster, M{\"u}nster D-48149, Germany}
\email{rmoll@soe.ucsc.edu}

\begin{abstract}
Oscillatory double-diffusive convection (ODDC) (also known as semi-convection) refers to a type of double diffusive instability that occurs in regions of planetary and stellar interiors which have a destabilizing thermal stratification and a stabilizing mean molecular weight stratification. In this series of papers, we use an extensive suite of three-dimensional (3D) numerical simulations to quantify the transport of heat and chemical species by ODDC. \citet{rosenblum2011} first showed that ODDC can either spontaneously form layers, which significantly enhance the transport of heat and chemical species compared to microscopic transport, or remain in a state dominated by large scale gravity waves, in which there is a more modest enhancement of the turbulent transport rates. Subsequent studies in this series have focused on identifying under what conditions layers form \citep{Mirouh2012}, and quantifying transport through layered systems \citep{Wood2013}. Here we proceed to characterize transport through systems that are unstable to the ODDC instability, but do not undergo spontaneous layer formation. We measure the thermal and compositional fluxes in non-layered ODDC from both 2D and 3D numerical simulations and show that 3D simulations are well approximated by similar simulations in a 2D domain. We find that the turbulent mixing rate in this regime is weak and can, to a first level approximation, be neglected. We conclude by summarizing the findings of papers I through III into a single prescription for transport by ODDC. 
\end{abstract}

\section{Introduction}
\subsection{Background}
Stratified fluids which have a stabilizing compositional gradient (Ledoux stable) and a destabilizing thermal gradient (Schwarzschild unstable) are common in a variety of astrophysical objects. This was first discussed by \citet{schwartzchildharm1958} in the context of the growing convection zones of massive stars $\left(>10M_{sun}\right)$. However, \citet{walin1964} and \citet{kato1966} were the first to realize that fluid instabilities may develop in such an environment, driving the turbulent transport of heat and chemical species. They are now commonly referred to as oscillatory diffusive instabilities, and they saturate into a non-linear weakly turbulent regime called Oscillatory Double Diffusive convection (ODDC). This nomenclature derives from the oscillatory nature of the basic unstable fluid motions which bear resemblance to over-stable internal gravity waves \citep{kato1966}. The efficiency of mixing (of heat and chemical species) in ODDC is potentially significant to evolution models for stars \citep{langer1985, merryfield1995} and giant planets \citep{stevenson1982}.

The conditions for ODDC to occur are defined by three non-dimensional parameters \citep{baines1969}. The Prandtl number, ${\rm Pr}$, and the inverse Lewis number, $\tau$, are defined as
\begin{equation}
\begin{array}{cccc}
{\rm Pr=\frac{\nu}{\kappa_{T}}} & , & \tau=\frac{\kappa_{\mu}}{\kappa_{T}} & ,
\end{array}
\end{equation}
where $\nu$ is the viscosity, and $\kappa_{T}$ and $\kappa_{\mu}$ are the thermal and compositional diffusivities, respectively. The third parameter, the inverse density ratio, is defined as
\begin{equation}
R_{0}^{-1} = \frac{\frac{\phi}{\delta}\nabla_{\mu}}{\nabla-\nabla_{\rm ad}} = \frac{ \frac{\phi}{\delta}\left( \frac{d \ln \mu}{d \ln p} \right) }{ \left( \frac{d \ln T}{d \ln p} \right) - \left( \frac{d \ln T}{d \ln p} \right)_{\rm ad} } \, ,
\end{equation}
where $T$, $p$ and $\mu$ are the temperature, pressure and mean molecular weight, and where $\delta = \left. \frac{\partial \ln \rho}{\partial \ln T} \right|_{p,\mu}$ and $\phi = \left. \frac{\partial \ln \rho}{\partial \ln \mu} \right|_{p,T}$ are thermodynamic derivatives of the equation of state. The subscript ``ad" is used to denote an adiabatic gradient (a derivative at constant entropy). ODDC occurs when $R_0^{-1}$ is within the following range \citep{kato1966, walin1964},
\begin{equation} 
1<R_{0}^{-1}<R_{\rm c}^{-1} \equiv \frac{{\rm Pr+1}}{{\rm Pr+\tau}} \, .
\end{equation}
When $R_{0}^{-1}$ is less than one, the system is unstable to overturning convection, and when $R_{0}^{-1}$ is greater than $R_{c}^{-1}$ the system is completely stable. A useful proxy for the inverse density ratio is the so-called ``reduced inverse density ratio" parameter \citep{Mirouh2012}, defined as,
\begin{equation} \label{eq:reducedR}
r=\frac{R_{0}^{-1}-1}{\frac{{\rm Pr+1}}{{\rm Pr+\tau}}-1} \, .
\end{equation}
This parameter maps the entire ODDC range to the interval $\left[0,1\right]$ where $0$ marks the onset of overturning convection, and $1$ marks the boundary between the ODDC parameter space and that of complete stability.

\subsection{Recent studies and current work}
In previous papers of this series we have set out to systematically characterize the different types of behavior demonstrated in ODDC. By analyzing 3D direct numerical simulations, \citet{rosenblum2011} first identified two general classes of behavior: one in which thermo-compositional layers emerge after the growth and saturation of the primary ODDC instability (and cause a significant increase in transport), and one in which layers do not form and where the dynamics are dominated by what they called ``homogeneous turbulence". \citet{Mirouh2012} built on the work of \citet{rosenblum2011} by developing a semi-analytical rule for identifying the regions of parameter space where layers do and do not form. They also determined that non-layered ODDC is ultimately dominated by large-scale internal gravity waves which (surprisingly) also augment thermal and compositional transport, though not as much as in the layered case. \citet{Wood2013} then studied the dynamics and transport properties of layered ODDC. In this work we now focus on characterizing the behavior of non-layered ODDC and quantifying the associated thermo-compositional mixing it induces.

In Section \ref{sec:MathMod} we present our mathematical model for the dynamics of ODDC. We also briefly describe how we study it using DNS, and discuss the metrics by which we measure thermal and compositional fluxes. In Section \ref{sec:bahavODDC}, we describe the evolution of a typical non-layered simulation and discuss the difficulties of running numerical simulations in extreme parameter regimes. In Section \ref{sec:2Dsim}, we discuss the effectiveness of 2D simulations for modeling ODDC compared to the full 3D DNS. We present the results of our numerical experiments in 2D and 3D, focussing in particular on the measurements of thermal and compositional fluxes in Section \ref{sec:results}. Finally, in Section \ref{sec:conclusion}, we present our conclusions, and summarize the findings of papers I, II, and III of this series.

\section{Mathematical model and numerical simulations} \label{sec:MathMod}
In what follows, we consider a region that is significantly smaller than the density scale height of a typical star or planet, and where flow speeds are much smaller than the sound speed of the medium. This allows us to use the Boussinesq approximation \citep{spiegelveronis1960} and to ignore the spherical geometry of an actual stellar or planetary interior. We consider a Cartesian domain where gravity is oriented in the negative $z$-direction, and we assume constant background gradients of temperature and mean molecular weight that are defined as follows,
\begin{eqnarray}
T_{0z} = \frac{dT}{dz} = \frac{T}{p}\frac{dp}{dz}\nabla \, , \nonumber \\
\mu_{0z} = \frac{\partial \mu}{\partial z} = \frac{\mu}{p}\frac{\partial p}{\partial z}\nabla_{\mu} \, .
\end{eqnarray}
We use the following linearized equation of state,
\begin{equation}
\frac{\tilde{\rho}}{\rho_0} = -\alpha \tilde{T} + \beta \tilde{\mu} \, ,
\end{equation}
where $\tilde{\rho}$, $\tilde{T}$ and $\tilde{\mu}$ are dimensional perturbations to the background profiles of density, temperature and composition, and where $\rho_0$ is the mean density of the region. The parameters $\alpha$ and $\beta$ are the coefficient of thermal expansion and coefficient of compositional contraction, respectively, and are defined as,
\begin{eqnarray}
\alpha = -\frac{\delta}{T} &= &-\left. \frac{1}{\rho_{0}}\frac{\partial\rho}{\partial T} \right|_{p,\mu} \, , \nonumber \\
\beta = \frac{\phi}{\mu} &= &\left. \frac{1}{\rho_{0}}\frac{\partial\rho}{\partial\mu} \right|_{p,T} \, .
\end{eqnarray}
Using these assumptions, the standard non-dimensional governing equations for ODDC \citep{rosenblum2011, Mirouh2012} are,
\begin{eqnarray} \label{eq:GovEqns}
\nabla\cdot\tilde{\textbf{u}} &= &0 \nonumber \, , \\
\frac{1}{{\rm Pr}}\left(\frac{\partial\tilde{\textbf{u}}}{\partial t}+\tilde{\textbf{u}}\cdot\nabla\tilde{\textbf{u}}\right) &= &-\nabla\tilde{p}+(\tilde{T}-\tilde{\mu})\textbf{e}_{z}+\nabla^{2}\tilde{\textbf{u}} \, , \nonumber \\
\frac{\partial\tilde{T}}{\partial t}+\tilde{\textbf{u}}\cdot\nabla\tilde{T}-\tilde{w} &= &\nabla^{2}\tilde{T} \, , \nonumber \\
\frac{\partial\tilde{\mu}}{\partial t}+\tilde{\textbf{u}}\cdot\nabla\tilde{\mu}-R_{0}^{-1}\tilde{w} &= &\tau\nabla^{2}\tilde{\mu} \, ,
\end{eqnarray}
where $\tilde{\mathbf{u}}$ is the velocity field, and all values with tildes are now non-dimensional. To arrive at these equations, we define the unit length in terms of (dimensional) constants,
\begin{equation}
\left[l\right] = d = \left(\frac{\kappa_{T}\nu}{\alpha g|T_{0z}-T_{0z}^{\rm ad}|}\right)^{\frac{1}{4}} \, ,
\end{equation} 
where $g$ is the gravitational acceleration, and where the parameter $T_{0z}^{\rm ad}$ is the background adiabatic temperature gradient which is defined as,
\begin{equation}
T_{0z}^{\rm ad} = \frac{T}{p}\frac{dp}{dz}\nabla_{\rm ad} \, .
\end{equation}

Using these constants, we also construct the units of time, $\left[ t \right]$, temperature, $\left[ \tilde{T} \right]$, and mean molecular weight, $\left[ \tilde{\mu} \right]$, as
\begin{eqnarray}
\left[t\right] &= &\frac{d^{2}}{\kappa_{T}} \, , \nonumber \\
\left[\tilde{T}\right] &= &d|T_{0z}-T_{0z}^{\rm ad}| \, , \nonumber \\
\left[\tilde{\mu}\right] &= &\frac{\alpha}{\beta}d|T_{0z}-T_{0z}^{\rm ad}| \, .
\end{eqnarray}

Furthermore, we can also rewrite the inverse density ratio $R_0^{-1}$ in terms of the parameters introduced above as,
\begin{equation}
R_{0}^{-1}=\frac{\beta|\mu_{0z}|}{\alpha|T_{0z}-T_{0z}^{\rm ad}|} \, .
\end{equation} 

As the physical characteristics of stellar and planetary interiors cannot be easily reproduced in laboratory experiments, we must rely on numerical simulations to gain insight about the nonlinear behavior of ODDC in these objects. We solve the set of equations in (\ref{eq:GovEqns}) using a pseudo-spectral code \citep{traxler2011} where all perturbations are triply periodic in the domain. Much of the experimental data presented in subsequent sections was generated by \citet{rosenblum2011} and \citet{Mirouh2012} in simulation runs using this code. We have generated new data specifically for this work as well, including a series of 2D simulations which we will compare to the 3D ones.

The main quantities of interest we extract from these numerical simulations are the turbulent vertical fluxes of temperature and chemical species, $\langle \tilde{w} \tilde{T} \rangle$ and $\langle \tilde{w} \tilde{\mu} \rangle$, respectively, where the angle brackets represent a spatial integral over the entire computational domain. It is often useful to express these fluxes in terms of thermal and compositional Nusselt numbers ($\rm Nu_T$ and $\rm Nu_{\mu}$) which are the ratios of the total fluxes to the diffusive fluxes (of temperature and composition). The turbulent fluxes are defined as follows in terms of $\rm Nu_T$ and $\rm Nu_{\mu}$,
\begin{eqnarray} \label{eq:TurbFluxes}
F_{T}^{\rm turb} &= &\langle \tilde{w} \tilde{T} \rangle ={\rm Nu}_{T}-1 \, , \nonumber \\
F_{\mu}^{\rm turb} &= &\langle \tilde{w} \tilde{\mu} \rangle =\tau R_{0}^{-1}\left({\rm Nu}_{\mu}-1\right) \, .
\end{eqnarray}

As shown by \citet{Wood2013}, these fluxes often exhibit fast oscillations with large amplitudes due to the gravity waves, so for the purposes of analysis, it can be more useful to consider related quantities known as the thermal and compositional dissipations, $\langle |\nabla \tilde{T}|^2 \rangle$ and $\langle |\nabla \tilde{\mu}|^2 \rangle$. Indeed, taking a spatial integral of the thermal and chemical evolution equations, and then assuming that the system is in a statistically stationary state, we find that the dissipations are related to the turbulent fluxes and Nusselt numbers by,
\begin{eqnarray} \label{eq:NusFluxDiss}
\overline{{\rm Nu }_T} - 1 = \overline{\langle \tilde{w} \tilde{T} \rangle} &= & \overline{\langle |\nabla \tilde{T}|^2 \rangle} \, , \nonumber \\
\overline{{\rm Nu}_{\mu}} - 1 = \frac{\overline{\langle \tilde{w} \tilde{\mu} \rangle}}{\tau R_{0}^{-1}} &= & \frac{\overline{\left<|\nabla \tilde{\mu}|^2\right>}}{\left(R_{0}^{-1}\right)^2} \, ,
\end{eqnarray}
where the bars indicate temporal averages over the entire statistically stationary period. In practice, these are good approximations even when the temporal averaging is done over short periods, so in what follows we assume similar relations between the \text{instantaneous} Nusselt numbers and dissipations as well. This is advantageous because the dissipations are less sensitive to the oscillations of gravity waves than the fluxes, and are therefore easier to analyze.

\section{Behaviors of ODDC} \label{sec:bahavODDC}
In what follows, we show the results of a typical ODDC simulation. The simulation presented has ${\rm Pr}=\tau=0.03$ and $R_{0}^{-1}=7.87$, and was run at an effective resolution of $192^3$ (the simulation domain has dimensions $\left( 100d \right)^3$). We first describe these results purely qualitatively, then move on to a more quantitative analysis.
\subsection{Qualitative study} \label{sec:qualBehav}
\begin{figure} 
\centering
\subfloat[]{\includegraphics[scale=0.25]{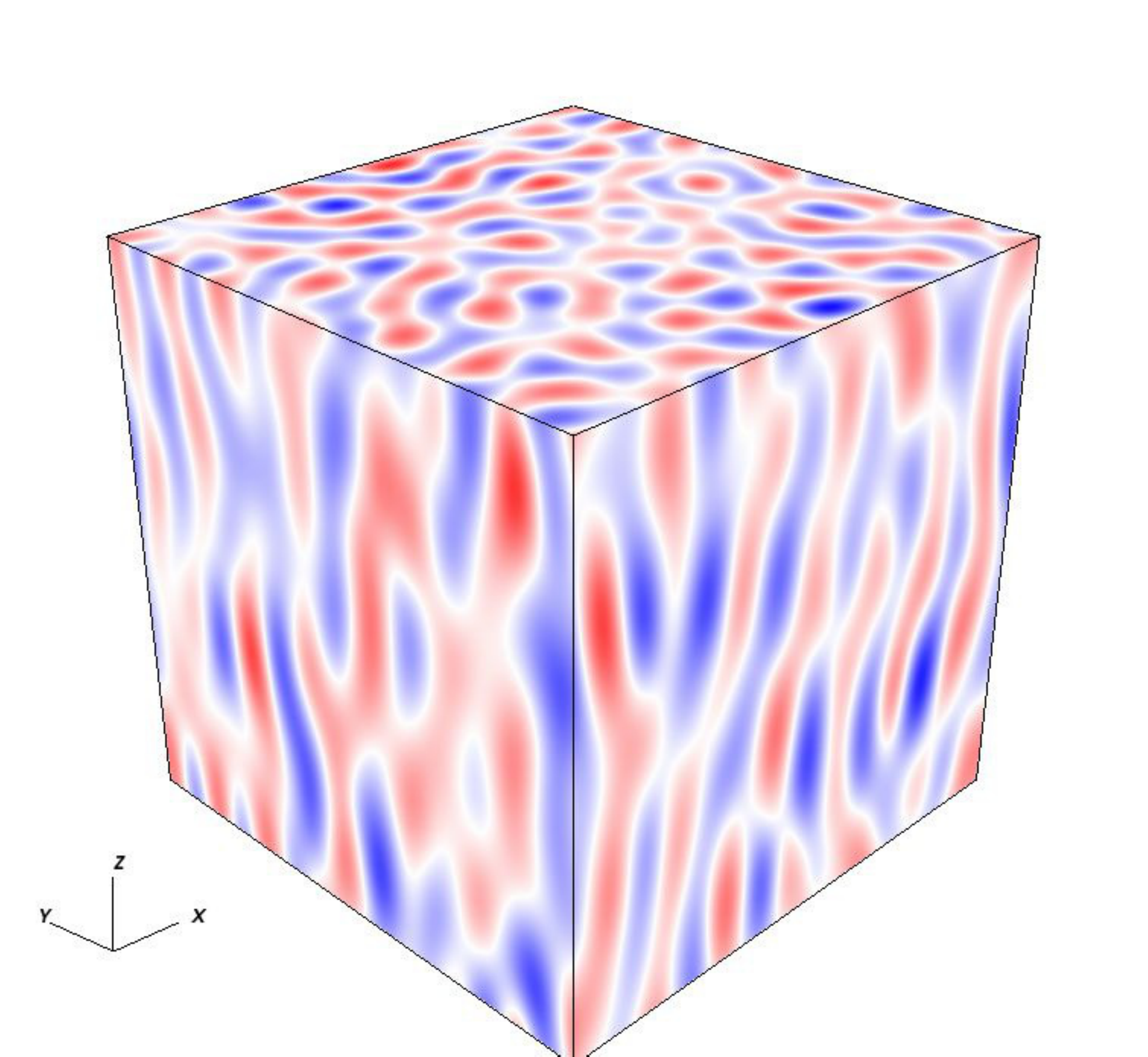} \label{fig:Early_a}} \hspace{1em}
\subfloat[]{\includegraphics[scale=0.25]{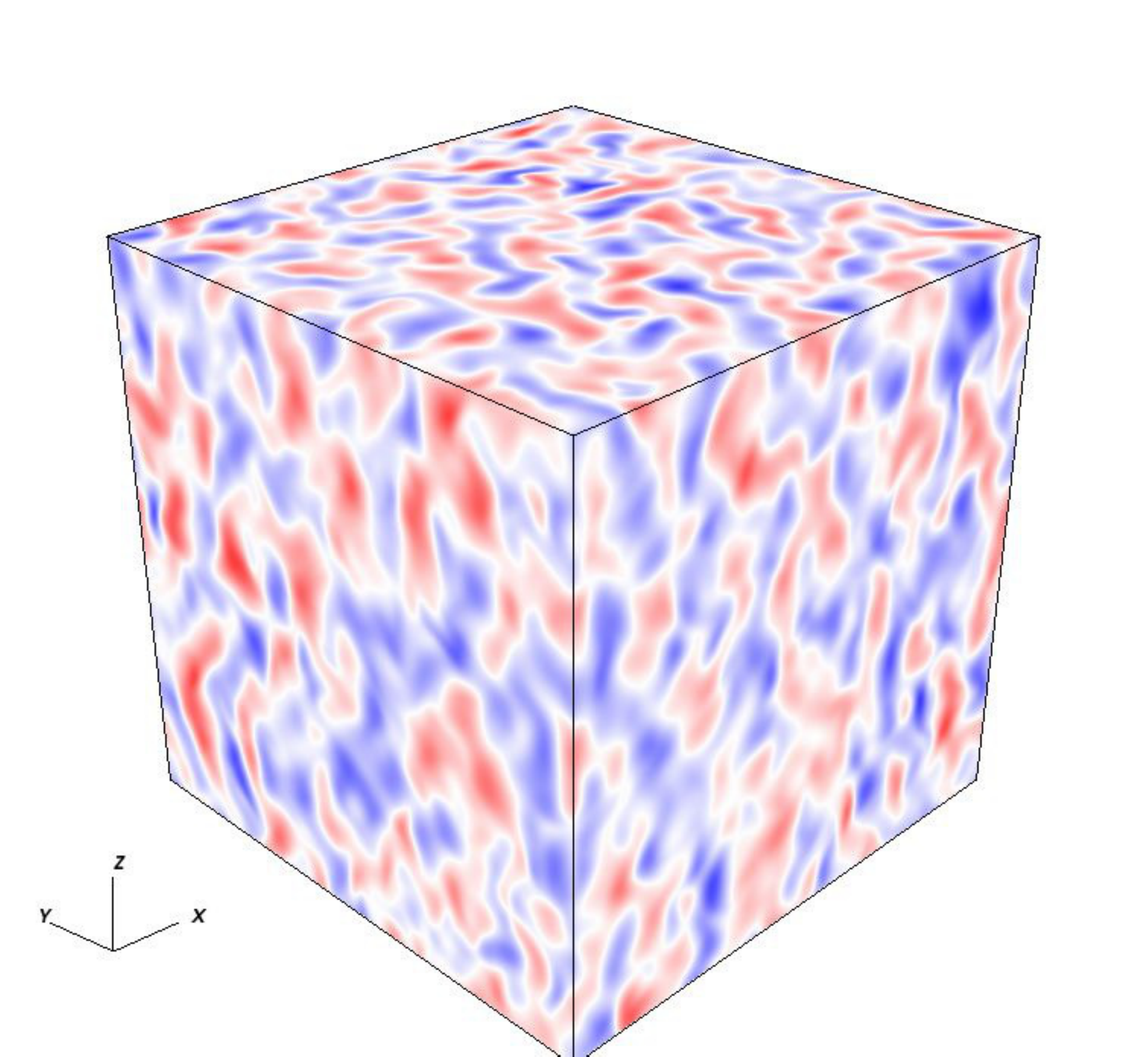} \label{fig:Sat_b}}
\caption{Figure \ref{fig:Early_a} shows the vertical component of velocity field during the basic instability growth phase $\left( t=2508 \right)$, with red and blue signifying upward and downward motion, respectively. Figure \ref{fig:Sat_b} shows the vertical component of the velocity field at the saturation of the primary instability $\left( t=2868 \right)$. For each figure, $R_0^{-1}=7.87$ and ${\rm Pr}=\tau=0.03$. \label{fig:sim_snapshots_early}}
\end{figure}

\begin{figure} 
\centering
\subfloat[]{\includegraphics[scale=0.25]{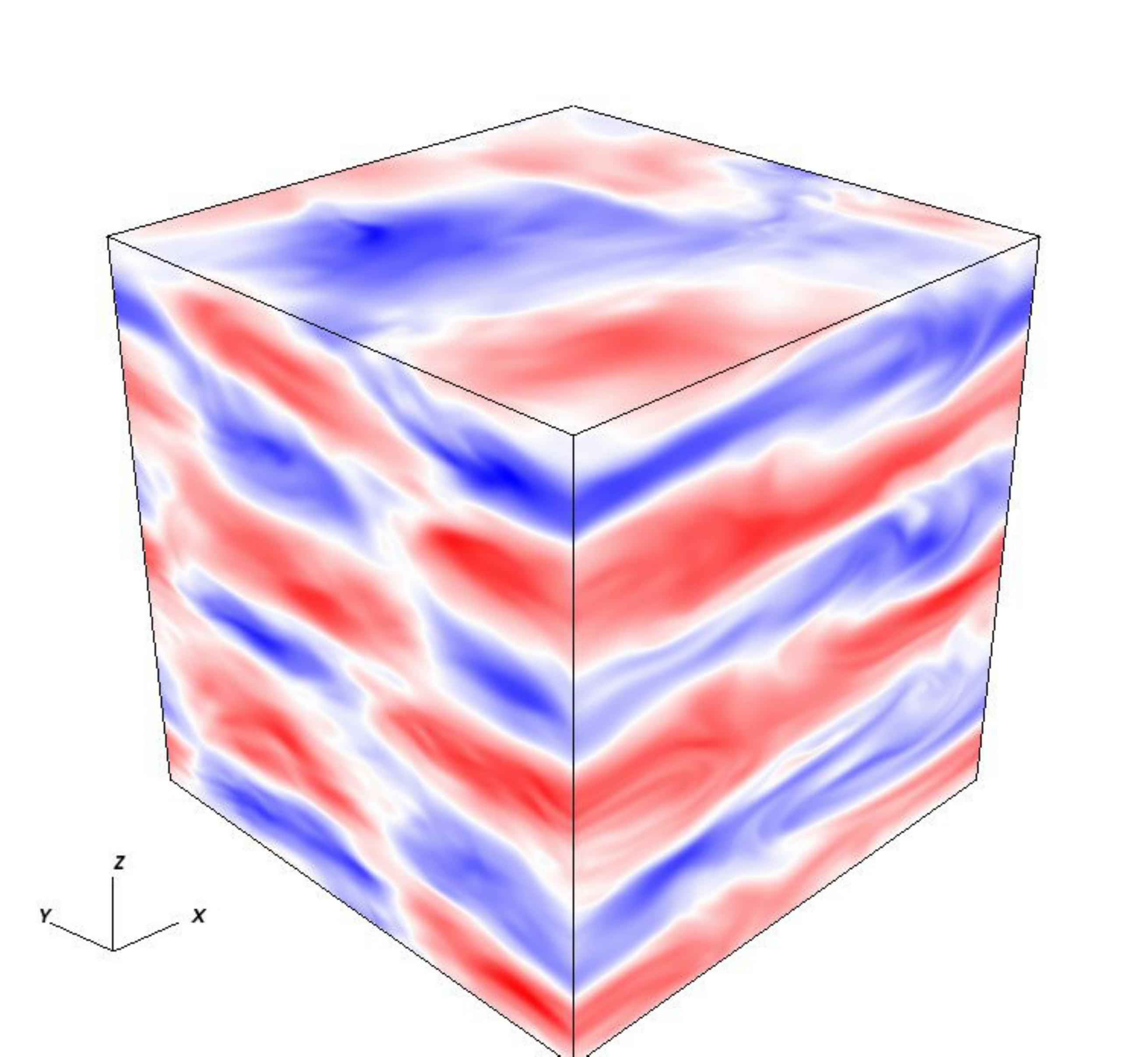} \label{fig:GrvWv_a}} \hspace{1em}
\subfloat[]{\includegraphics[scale=0.25]{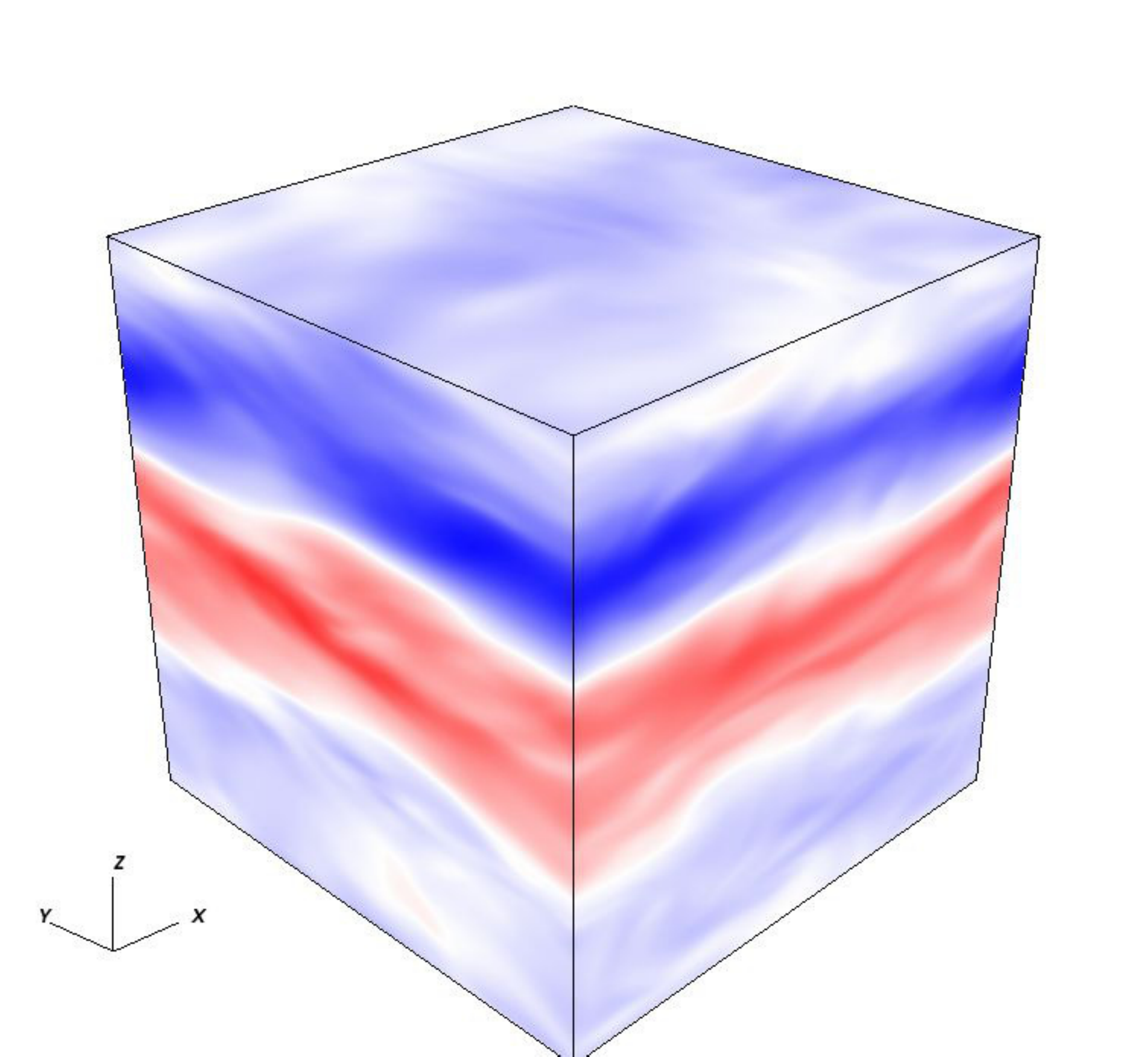} \label{fig:Shear_b}}
\caption{The Figures \ref{fig:GrvWv_a} and \ref{fig:Shear_b} show, respectively, the velocity field in the $y$ direction at times when the system is dominated by gravity waves $\left( t=20758 \right)$ and when the system is dominated by shear $\left( t=24718 \right)$. Here red and blue represent motion in the positive and negative $y$ direction. As with the snapshots in Figure \ref{fig:sim_snapshots_early}, $R_0^{-1}=7.87$ and ${\rm Pr}=\tau=0.03$. \label{fig:sim_snapshots_late}}
\end{figure}

Figure \ref{fig:sim_snapshots_early} shows the vertical component of the velocity field at early times, and Figure \ref{fig:sim_snapshots_late} shows the $y$-component of the velocity field later on in the simulation. At very early times, we first see the development of the fastest growing modes of the basic ODDC instability, which resemble tubes of vertically oscillating fluid (shown in Figure \ref{fig:Early_a}). This primary growth phase ends when the basic instability saturates due to nonlinear interactions inherent to the problem (see Figures \ref{fig:Sat_b}).

After the primary saturation, the small-scale fastest growing modes of the primary instability stop growing, but other larger-scale modes slowly continue to gain energy (amounting to a secondary phase of growth). From Figure \ref{fig:GrvWv_a} we see that the latter (which ultimately come to dominate the energetics of the system) generally have the largest possible scale in the horizontal direction. As the system evolves, energy is funneled into modes of progressively larger vertical scale until the secondary growth phase saturates and the system reaches a statistically stationary state. The dynamics of this state are characterized by gravity wave oscillations on fast timescales whose amplitudes are modulated chaotically and intermittently. This intermittency appears to be caused by nonlinear interactions between large scale gravity modes and large scale horizontal shearing modes. Indeed, we regularly observe the emergence of a strong horizontal shear layer that temporarily suppresses the wave field (Figure \ref{fig:Shear_b}). The shear then dissipates, and the system proceeds as before. Figure \ref{fig:sim_snapshots_late} shows the distinct differences in the $y$ velocity field between a gravity-wave-dominated phase and a shear-dominated phase.

\subsection{Quantitative study} \label{sec:quantBehav}
In order to study this system in a more quantitative way we now investigate the energy contained in individual modes. We shall refer to specific spatial modes by the number of wavelengths in the $x$, $y$, and $z$ directions. So mode $\left(l,m,k\right)$ would refer to a mode with horizontal wave numbers $\frac{2\pi l}{L_x}$ and $\frac{2\pi m}{L_y}$ (in the $x$ and $y$ directions), and vertical wave number $\frac{2\pi k}{L_z}$ (in the $z$ direction). We can quantify the transfer of energy to larger scales by considering the amount of energy in a mode ``family". A family of modes consists of all the modes with equivalent spatial structures given the symmetries between the $x$ and $y$ directions in the domain, and negative and positive wave numbers in each spatial direction. For example, the mode family 102 contains the modes (1,0,2) , (-1,0,2) , (0,1,2) , (0,-1,2) , (1,0,-2) , (-1,0,-2) , (0,1,-2) , (0,-1,-2) \citep{traxler2011}.

\begin{figure} 
\centering
\epsscale{0.9}
\plotone{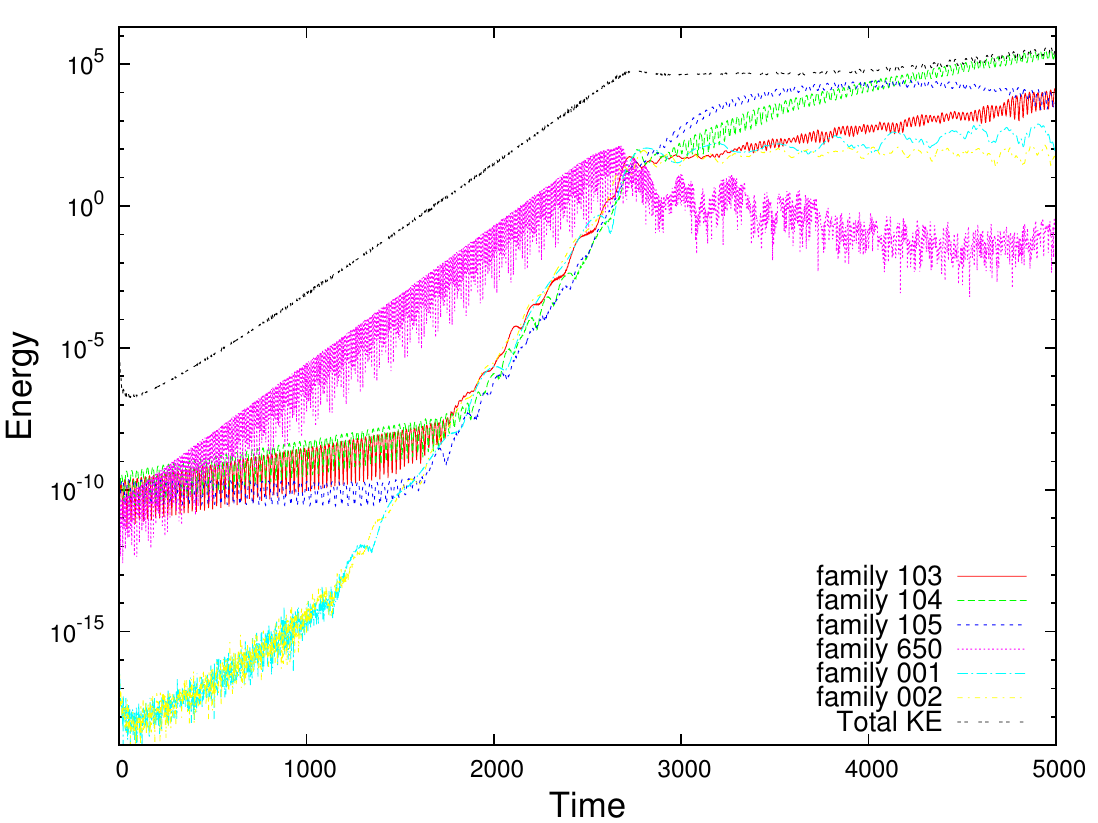}
\caption{Total kinetic energy vs. time for various families of modes (see main text for detail) for a simulation with ${\rm Pr}=\tau=0.03$ and $R_{0}^{-1}=7.87$. Shown is the early part of a simulation where the total kinetic energy is dominated by modes that are predicted to grow the fastest according to linear theory. The mode family (6,5,0) is one of the fastest growing mode families. \label{fig:mode-by-mode-early}}
\end{figure}

\begin{figure} 
\centering
\epsscale{0.9}
\plotone{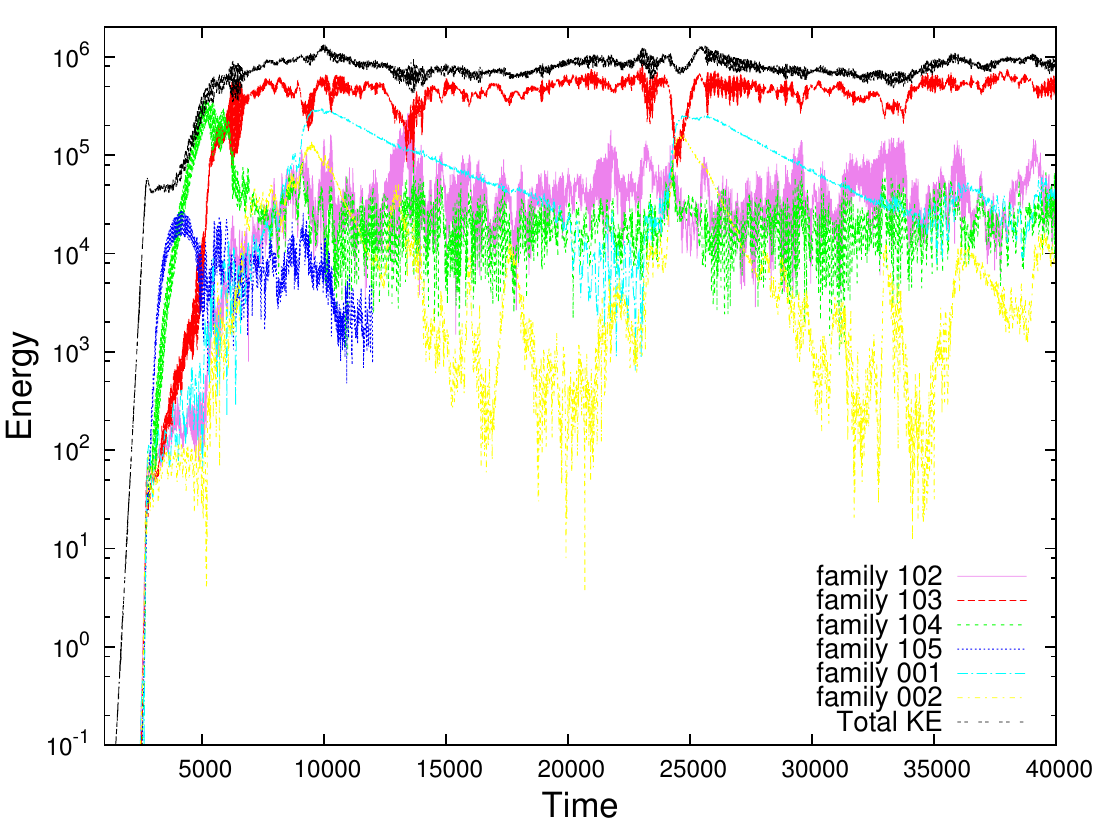}
\caption{As in Figure \ref{fig:mode-by-mode-early} (same parameters), but now focusing on the long-term evolution of the system. After the saturation of the primary instability, large scale gravity waves begin to grow and energy becomes more concentrated in larger scale modes as time goes on. Large scale shearing mode families (mode families (0,0,1) and (0,0,2)) also grow to large amplitude. \label{fig:mode-by-mode-full}}
\end{figure}

Consistent with the qualitative results in Figure \ref{fig:GrvWv_a}, Figure \ref{fig:mode-by-mode-early} shows that at early times, the dynamics of ODDC are dominated by horizontally small scale modes ($\sqrt{l^2+m^2} \approx 8$), with no structure in the vertical direction (with vertical wavenumber $k=0$). Around $t=2500$, the primary instability saturates. \citet{Mirouh2012} demonstrated that the level at which the primary instability saturates can be used to identify regions of parameter space where layer formation is expected to occur. However, the primary saturation level is of little use for characterizing the long term transport properties of our non-layered ODDC because a secondary growth phase occurs after primary saturation, augmenting the thermal and compositional fluxes. From Figure \ref{fig:mode-by-mode-full}, we see that while the fastest growing modes of the primary instability stop growing at saturation, the mode family (1,0,5) continues to grow, and for a brief time becomes the most energetic mode family in the system. As time goes on, however, the mode family (1,0,4) supplants (1,0,5) as the most energetic, which is in turn overtaken by the mode family (1,0,3). For this particular simulation, the handoff of energy to larger scale modes stops with mode family (1,0,3); mode family (1,0,2) never becomes dominant. Crucially, Figures \ref{fig:mode-by-mode-early} and \ref{fig:mode-by-mode-full} also reveal the growth of the energy in large scale shearing modes with purely horizontal fluid motions (mode families (0,0,1) and (0,0,2)). This is unexpected because these modes are not directly excited by the primary ODDC instability. Instead their growth must arise from nonlinear interactions between rapidly growing ODDC modes.

\begin{figure} 
\centering
\epsscale{0.65}
\plotone{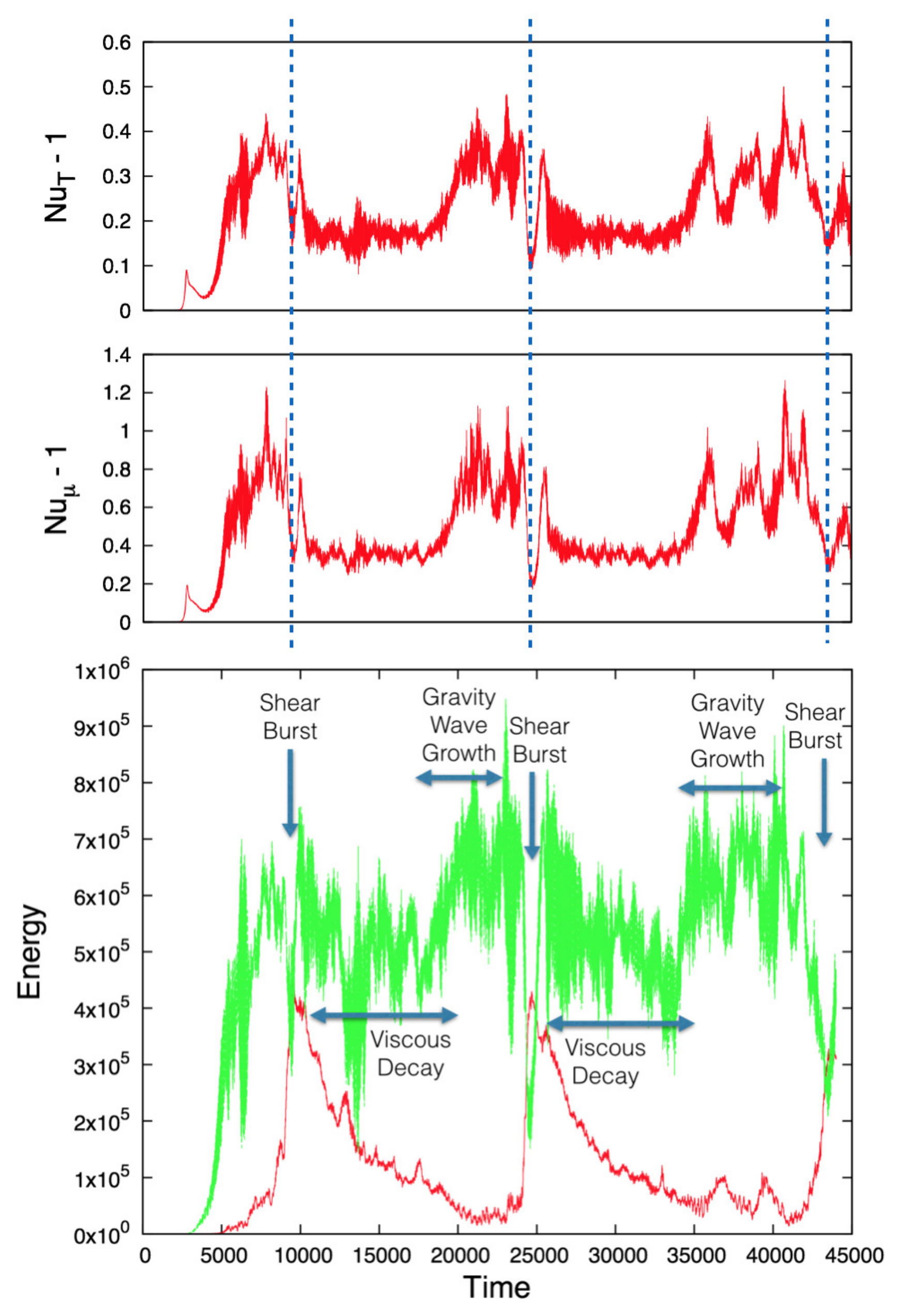}
\caption{The top figures show the fluxes of temperature and composition in terms of the Nusselt numbers, which are related to the turbulent fluxes by the equations in (\ref{eq:NusFluxDiss}). Shown in the bottom figure is the total kinetic energy in the large-scale gravity wave perturbations (modes from the families (1,0,n)) in green, and the kinetic energy in the background shear (modes from the families (0,0,m)) in red, as a function of time. The vertical lines mark times at which there are bursts of energy in the shear. The simulation parameters used here are ${\rm Pr} =\tau =0.03$, and $R_0^{-1}=7.87$. \label{fig:Pert_vs_horiz}}
\end{figure}

Around $t=5000$, after mode family (1,0,3) becomes dominant, the secondary growth phase appears to end, saturating into a statistically steady turbulent state. However, Figure \ref{fig:mode-by-mode-full} shows periodic bursts of growth in the shearing mode energies, suggesting intermittent (yet powerful) interactions between the shearing modes and the dominant gravity wave modes. In these interactions, illustrated in more detail in Figure \ref{fig:Pert_vs_horiz}, the growth of gravity waves excites the rapid growth of horizontal shearing modes, which in turn causes a decay in the amplitude of the large scale gravity waves. Without the wave field to amplify it, the shear then decays viscously. This finally allows the energy in the gravity waves to ramp back up again. While this interaction does not always manifest itself as such a distinct sawtooth oscillation, it is still present in one form or another in all gravity-wave-dominated ODDC simulations. Figure \ref{fig:Pert_vs_horiz} also shows that the interaction between the shear and gravity waves also leads to intermittent modulation of the thermal and compositional fluxes. 

While the intermittency in the fluxes caused by the shear is interesting in its own right, for the purpose of modeling ODDC transport in planetary or stellar evolution calculations, we are more concerned with estimating the mean fluxes over significant periods of time. These mean fluxes at secondary saturation depend on the parameters of the system ($R_{0}^{-1}$, $\rm Pr$, and $\tau$). The results shown in this section, which were obtained at moderate $R_{0}^{-1}$ and moderate $\rm{Pr}$ and $\tau$, suggest that turbulent transport in non-layered ODDC is weak. To confirm this we need to run numerical experiments at larger $R_{0}^{-1}$ and smaller $\rm{Pr}$ and $\tau$. Probing this region of parameter space is difficult, however, because 3D simulations at low $\rm{Pr}$ and $\tau$ can be computationally very expensive, particularly for values of $R_{0}^{-1}$ that are close to marginal stability $\left( R_0^{-1} \rightarrow \frac{{\rm Pr} + 1}{{\rm Pr} + \tau} \right)$. The small values of $\rm{Pr}$ and $\tau$ lead to small-scale turbulent features with steep gradients of velocity, temperature, and composition, which necessitate high spatial resolution. Furthermore, a larger $R_{0}^{-1}$ leads to higher frequency oscillations of the basic ODDC modes, necessitating smaller time steps. Given these challenges, in the next section we discuss the possibility of using 2D ODDC simulations as a potential surrogate for full 3D simulations at these extreme regions of parameter space.  

\section{2D simulations} \label{sec:2Dsim}
\begin{figure} 
\centering
\subfloat[]{\includegraphics[scale=0.75]{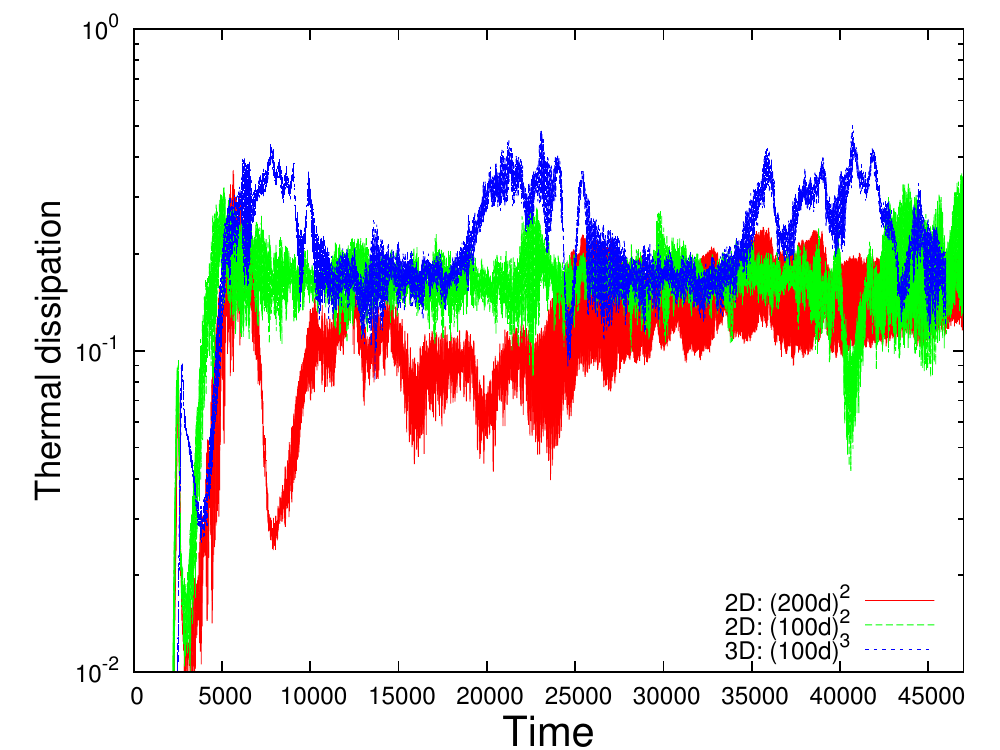} \label{fig:TCompare_a}} \hspace{1em}
\subfloat[]{\includegraphics[scale=0.75]{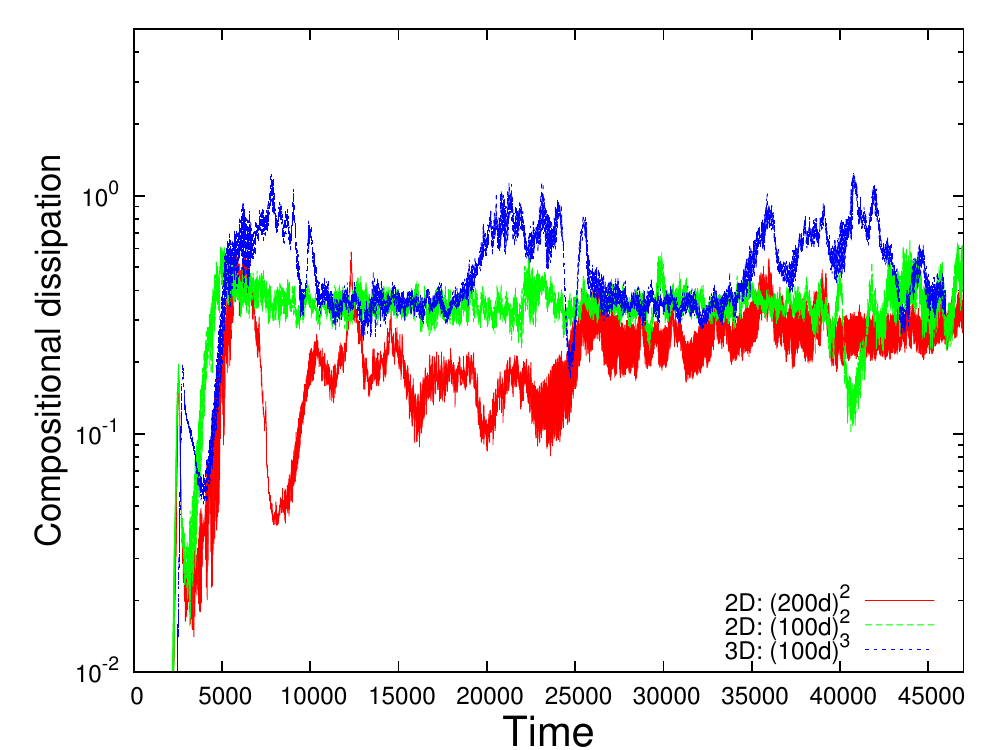} \label{fig:SCompare_b}}
\caption{Figures \ref{fig:TCompare_a} and \ref{fig:SCompare_b} show the thermal and compositional dissipations vs. time for the simulation with ${\rm Pr}=\tau=0.03$ and $R_{0}^{-1}=7.87$. Included are data from a 3D simulation, and two 2D simulations with differing domain sizes. \label{fig:2Dvs3D}}
\end{figure}

Simulations of 2D ODDC systems are computationally inexpensive and are also less intensive in terms of data storage than 3D simulations. For this reason, we have run a series of tests to compare both qualitative behavior and quantitative estimates of the fluxes (and other system diagnostics) in 2D and 3D. Fortunately, as we see from Figure \ref{fig:2Dvs3D}, the secondary saturation level in the 2D simulation at ${\rm Pr}=\tau=0.03$ and $R_0^{-1}=7.87$ is very similar to the 3D simulation analyzed in the previous section. This is, in fact, generally the case for each parameter regime where we have both 2D and 3D simulations, which is surprising in light of the fact that 2D simulations of overturning convection (i.e. $R_0^{-1} < 1$) behave very differently from their 3D counterparts \citep{schmalzl2004,vanderpoel2013}. In the results that follow in section \ref{sec:results}, we shall rely heavily on 2D simulations to draw conclusions about turbulent fluxes through non-layered ODDC.

\begin{figure} 
\centering
\subfloat[]{\includegraphics[scale=.75]{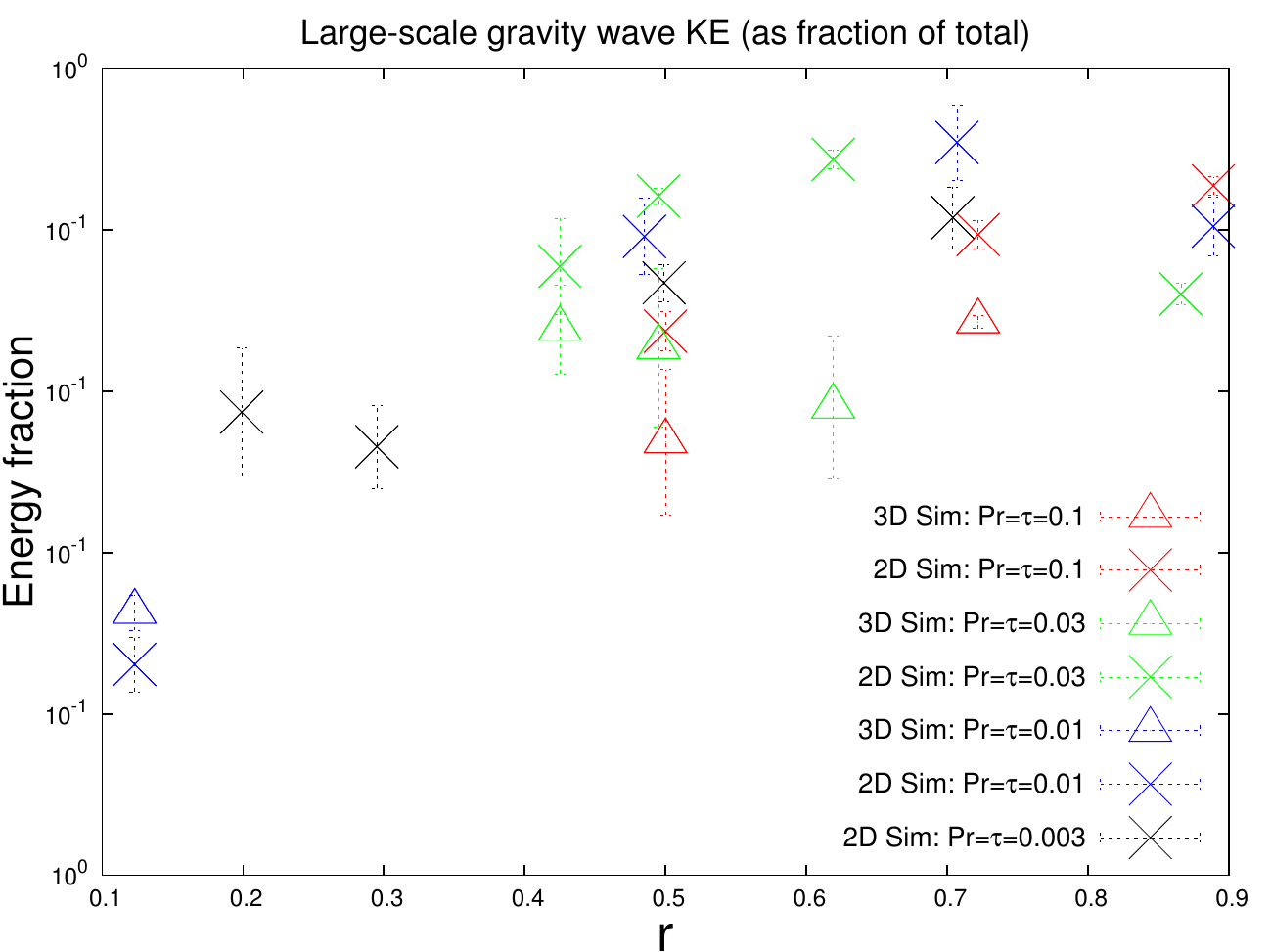} \label{fig:grvWvPct}} \\
\subfloat[]{\includegraphics[scale=.75]{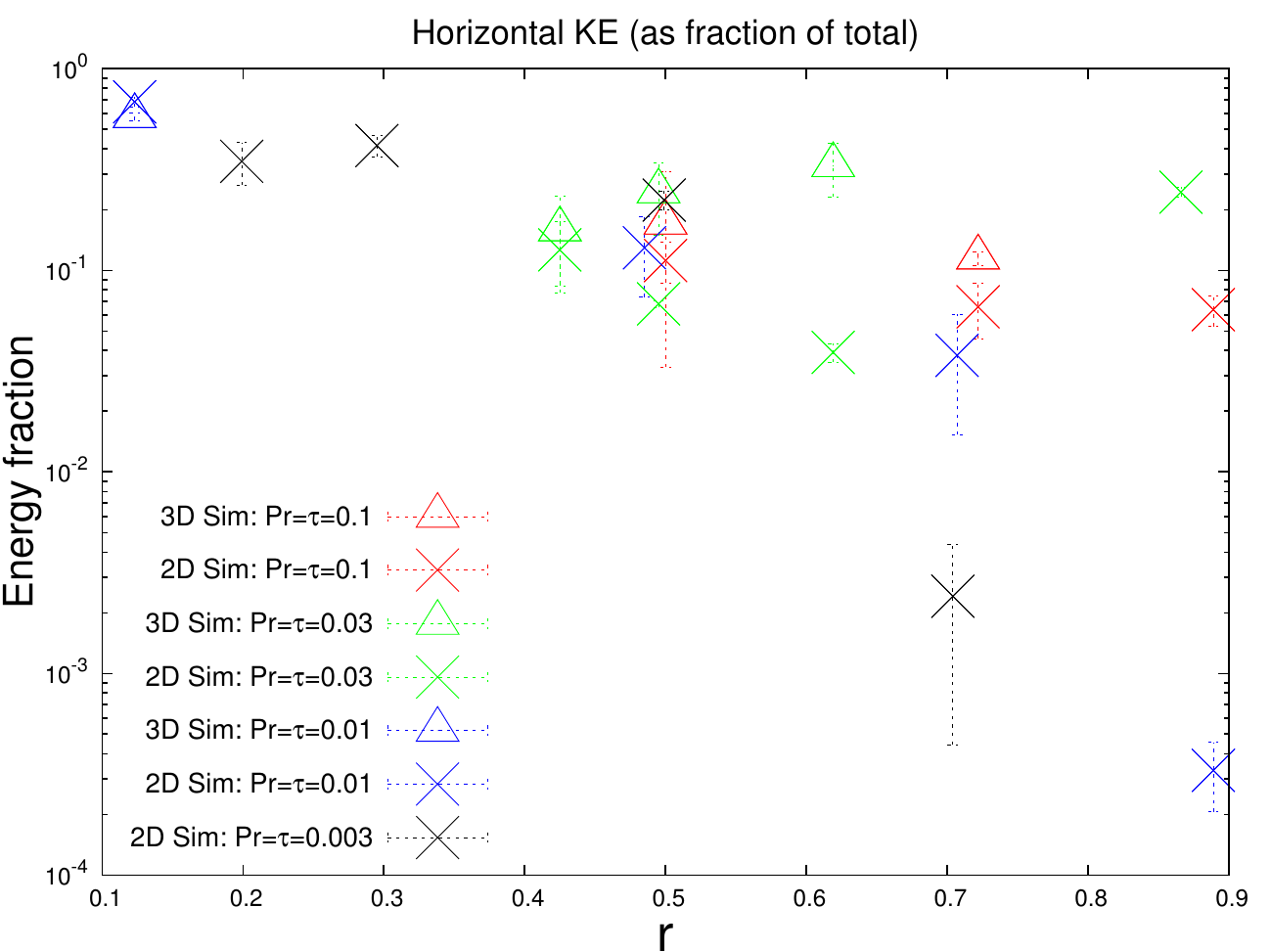} \label{fig:horizPct}}
\caption{Figure \ref{fig:grvWvPct} shows the energy in gravity wave families of the form (1,0,n) as a percentage of the total energy in the system. Figure \ref{fig:horizPct} shows the energy in shearing modes families of the form (0,0,m) as a percentage of the total energy. We estimate errors according to the method described in Section \ref{sec:results}. \label{fig:GrvWv_Shear_energy}}
\end{figure}

Measurements of mode family energies show that key physical processes that dictate the behavior of 3D ODDC simulations are present in the 2D simulations as well. Figure \ref{fig:GrvWv_Shear_energy} explores the energetics of the gravity waves and shear, and shows that the fractions of energy in each type of mode are of the same in order in both cases. This is important because together these two types of modes contain most of the energy in non-layered systems after secondary saturation.

The computational economy of 2D simulations makes other types of analysis easier as well, such as running simulations in larger domains. In the previous section, we showed that after primary saturation the dominant gravity wave modes have horizontal wavelengths commensurate with the domain size. Also we showed that energy is transferred to modes with progressively larger vertical wavelengths. This raises the question of whether this energy transfer would always terminate at a vertical wavelength that depends on the domain size. For example, will the dominant mode after secondary saturation in a $(200d)^3$ domain have a vertical wavelength that is twice that of the dominant mode in a $\left( 100d \right)^3$ domain? More importantly, do the fluxes depend on the domain size?

Using 2D data we find that in all but one case, doubling the domain size in each direction, leaves the vertical wavelength of the dominant mode unchanged. By contrast, the horizontal wavelength of the dominant mode always grows to the largest possible scale allowed by the domain. As a consequence, the dominant modes in the larger boxes are inclined more toward the horizontal than in the smaller ones. Importantly though, Figures \ref{fig:TCompare_a} and \ref{fig:SCompare_b} show that despite some qualitative differences between simulations with domains of different dimensions, we find that the time-averaged fluxes of temperature and chemical composition do not depend strongly on box size (they are within $\thicksim10\%$ of one another).

\section{Results and Discussion} \label{sec:results}
We now analyze the results of all numerical experiments done in the 2D and 3D computational domains. We evaluate thermal and compositional fluxes in terms of the Nusselt numbers, which we calculate from thermal and compositional dissipation data, as described in (\ref{eq:NusFluxDiss}). More specifically, the quantities we consider are ${\rm Nu}_T - 1$ and ${\rm Nu}_{\mu} - 1$ which, conveniently, are also measures of the turbulent fluxes in units of the diffusive fluxes. To find typical Nusselt numbers for a simulation we calculate time averaged values for the period after secondary saturation. We define the secondary saturation time to be the point at which the total kinetic energy of the system reaches a statistically stationary level, and we identified it by inspection. To estimate errors we first divide the time average domain into 10 bins. We then take the average of each bin and calculate the standard deviation of the individual bin averages from the overall average. 

\begin{figure} 
\centering
\subfloat[]{\includegraphics[scale=.75]{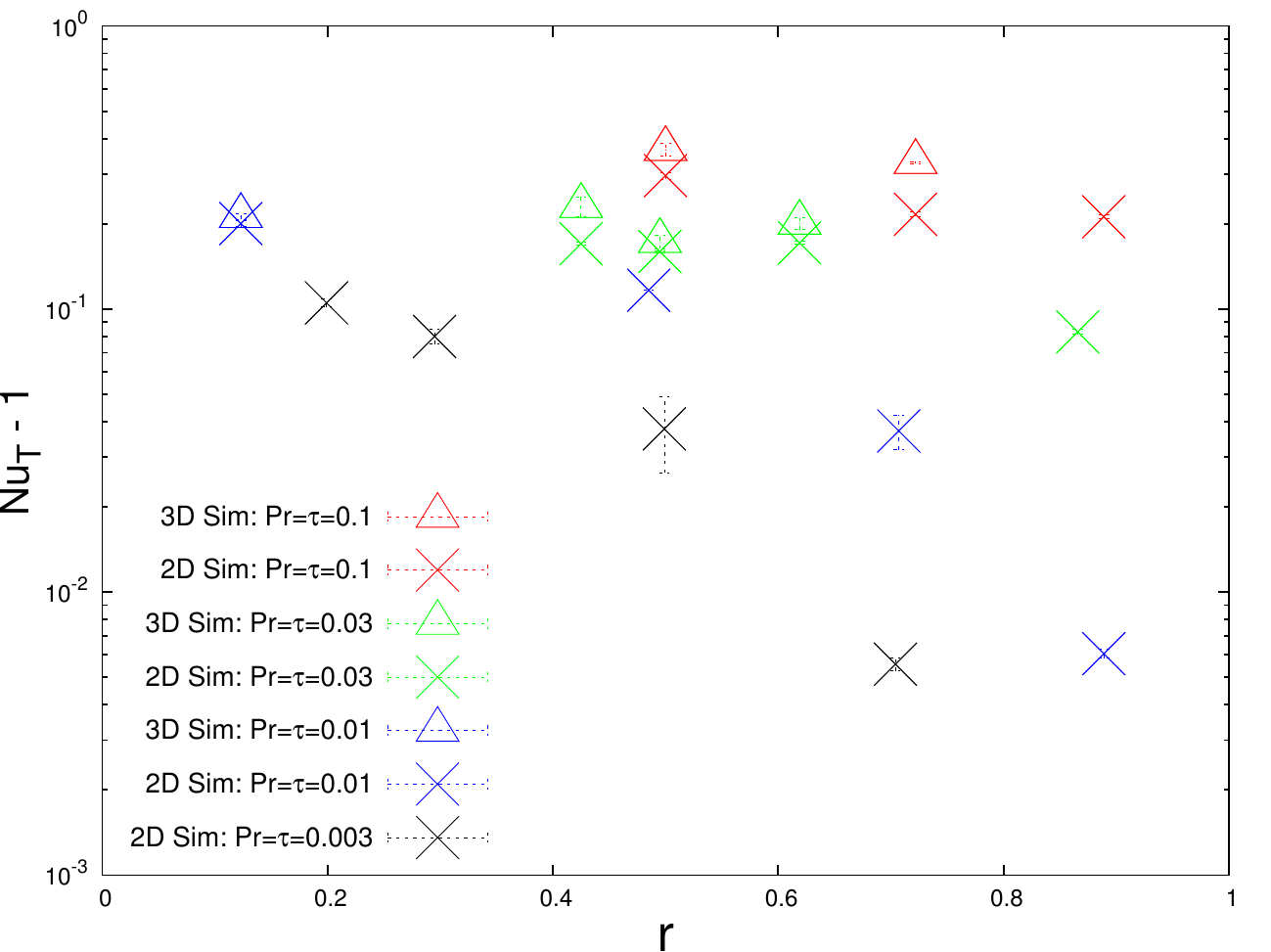} \label{fig:TSat_level}} \\
\subfloat[]{\includegraphics[scale=.75]{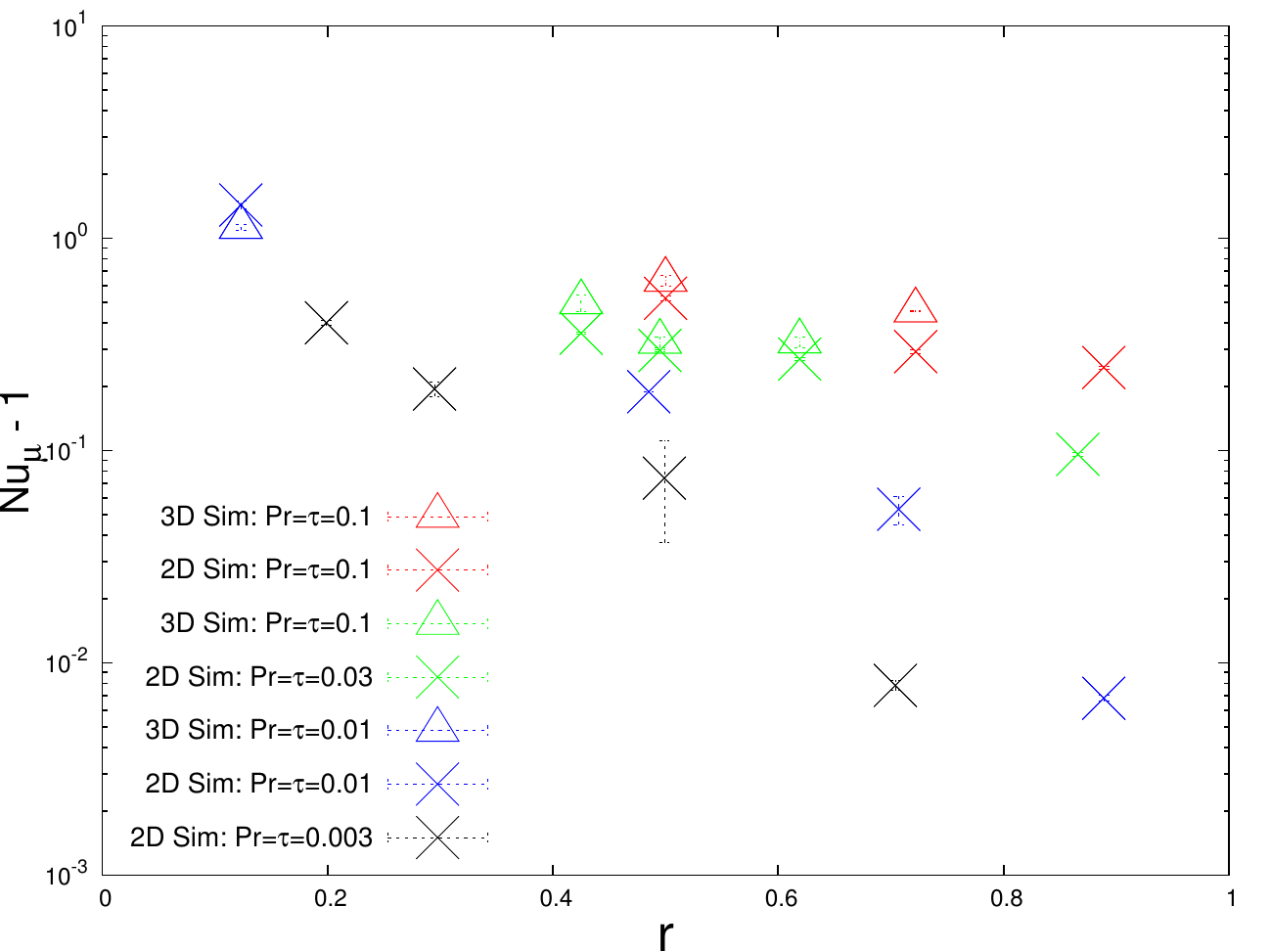} \label{fig:SSat_level}}
\caption{Figures \ref{fig:TSat_level} and \ref{fig:SSat_level} show the Nusselt numbers after secondary saturation for the available 3D and 2D simulations (in domains of size $100d^3$ or $100d^2$) as a function of $r$ (which is related to $R_0^{-1}$ by Equation \ref{eq:reducedR}). The secondary saturation level decreases as $r$ increases and as ${\rm Pr}$ and $\tau$ decrease. \label{fig:FluxComp}}
\end{figure}

Figure \ref{fig:FluxComp} shows the thermal and compositional Nusselt numbers calculated in both cases, for various values of $\rm{Pr}$ and $\tau$. From this figure we also see that the reasonable agreement between 2D and 3D simulations discussed in Section \ref{sec:2Dsim} persists in all cases studied. 

The primary result of this analysis is that, while gravity waves consistently augment thermal and compositional transport in non-layered ODDC, the enhanced fluxes are only slightly larger than the transport due to thermal and molecular diffusion alone. To be precise, the turbulent transport due to gravity waves decreases with increasing $R_0^{-1}$. For the simulations we have run with the smallest values of $R_{0}^{-1}$ (those closest to the layering threshold), the turbulent compositional flux is at most twice that of the flux due to diffusion alone and the turbulent heat flux is at most $\thicksim20\%$ of the diffusive flux. However, at larger values of $R_{0}^{-1}$, closer to marginal stability, the turbulent fluxes drop down to $\thicksim5-10\%$ of the diffusive fluxes. Also, critically, as ${\rm Pr}$ and $\tau$ decrease, the fluxes decrease as well. Consequently, the simulations run at the parameter regime most similar to actual giant planetary interiors (${\rm Pr}=\tau=0.003$) showed the smallest turbulent fluxes compared to other simulations with the same value of $r$, but with larger ${\rm Pr}$ and $\tau$. 

\begin{figure} 
\centering
\epsscale{0.65}
\plotone{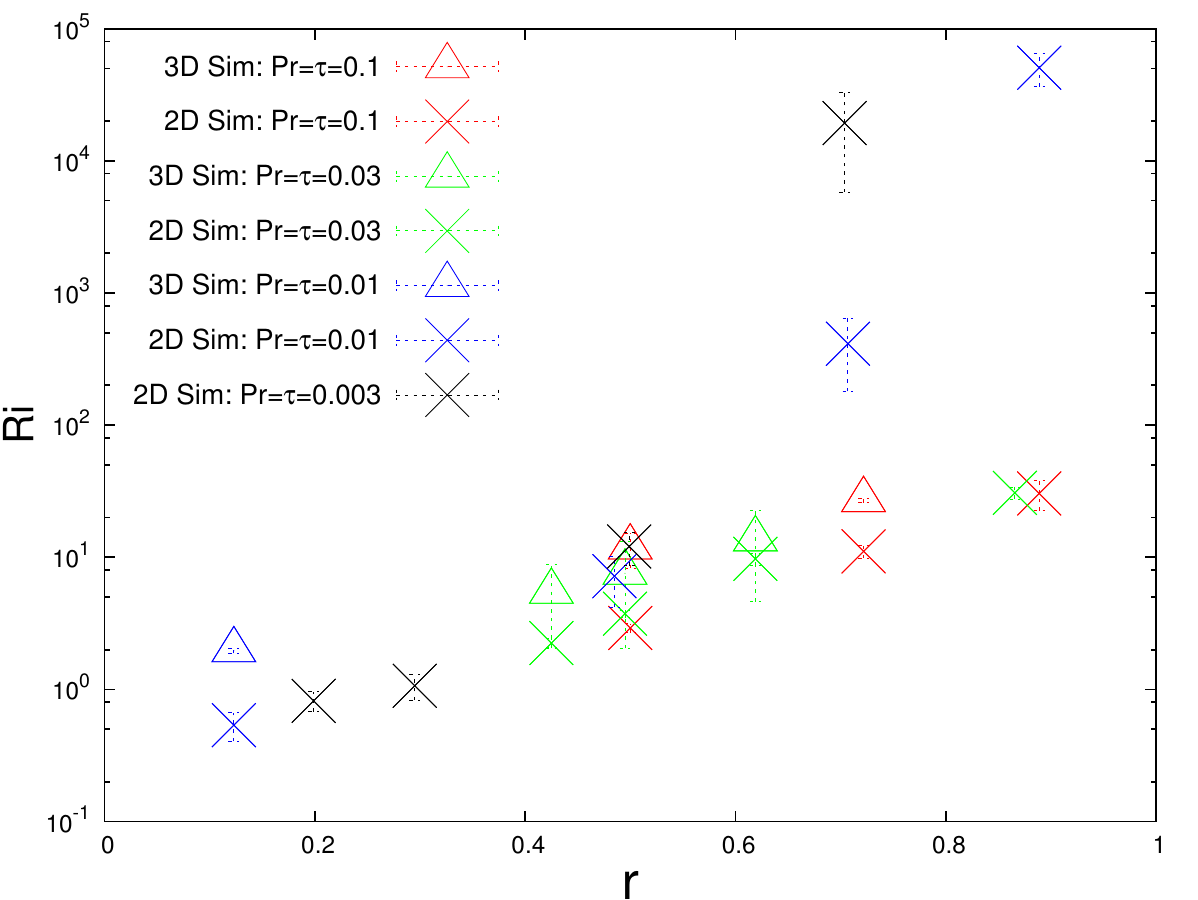}
\caption{The plot shows the time- and horizontally-averaged minimum Richardson number as a function of $R_0^{-1}$. Included are data from all available 2D and 3D simulations (in domains of size $100d^3$ or $100d^2$). \label{fig:Rich_Num}}
\end{figure}

Another result of our analysis is that gravity wave dominated ODDC is responsible for the generation of large scale shear. In all the simulations we have run so far, the main effect of the shear has been to modulate the wave-induced transport through strong non-linear interactions with the wave field. One might wonder, however, whether the shear could become strong enough in some parameter regimes to trigger a shear instability which would then dramatically augment turbulent transport. To evaluate the likelihood of this happening, we consider the Richardson number, ${\rm Ri}$, which is the ratio of the available kinetic energy in the shear to the potential energy needed to cause overturn. In the units of this paper, we define the Richardson number as,
\begin{equation}
{\rm Ri}\left(z\right) = \frac{N^2}{|\frac{\partial \mathbf{u}}{\partial z}|^2} \simeq {\rm Pr} \left( \frac{ R_0^{-1} - 1}{\left(\frac{d\bar{u}}{dz}\right)^2 + \left(\frac{d\bar{v}}{dz}\right)^2} \right) \, ,
\end{equation}
where $N$ is the buoyancy frequency, defined dimensionally as $N^2 = -g\frac{d \ln \rho}{dz}$ where $\rho$ is the background density profile. The terms $\bar{u}$ and $\bar{v}$ are the horizontally averaged $x$ and $y$ components of velocity, respectively (for 2D simulations $v=0$ everywhere, for all time). To calculate the typical minimum Richardson number for a simulation we find the minimum of ${\rm Ri}(z)$ for an individual time step, and then take a time average of ${\rm Ri}_{\rm min}(z)$ over the period after secondary saturation. A plot of the time-averaged minimum Richardson number of the available simulations (Figure \ref{fig:Rich_Num}) shows that ${\rm Ri}_{\rm min}$ increases as $r$ (or equivalently, $R_0^{-1}$) increases. This is due to the fact that by definition systems with lower $R_0^{-1}$ have a weaker stabilizing compositional stratification compared to their unstable thermal stratification, making them more susceptible to overturning. Also, recall from Figure \ref{fig:GrvWv_Shear_energy} that simulations with higher values of $R_0^{-1}$ have a lower percentage of their total kinetic energy is in shearing modes. This Richardson number data therefore suggests that if any shear-induced instabilities were to present themselves, they would do so at values of $R_0^{-1}$ that are closer to the layering threshold.

\section{Conclusion} \label{sec:conclusion}
This study marks the conclusion of a series of papers aimed at fully describing the thermal and compositional flux properties of ODDC, throughout the entire ODDC parameter regime. \citet{rosenblum2011} laid the groundwork for this series by conducting a general survey of the ODDC parameter space and identified that ODDC either spontaneously form layers, or remains in a non-layered state, which \citet{Mirouh2012} later showed to be dominated by large scale gravity waves. They also showed that the critical parameter to making predictions about layer formation is the inverse total flux ratio $\gamma_{tot}^{-1}$ defined as,  
\begin{equation}
\gamma_{\rm tot}^{-1} = \tau R_0^{-1} \frac{{\rm Nu}_{\mu}}{{\rm Nu}_T} = \frac{\tau R_0^{-1} + \gamma_{\rm turb}^{-1}\left({\rm Nu}_T-1\right)}{1+\left({\rm Nu}_T-1\right)} \, ,
\end{equation}
where $\gamma_{\rm turb}^{-1}$ is the inverse \textit{turbulent} flux ratio. More precisely, \citet{rosenblum2011} showed that \citep[similar to the conditions that lead to layer formation in fingering convection][]{radko2003mlf} layers only form when,
\begin{equation}
\frac{d \gamma_{\rm tot}^{-1}}{d R_0^{-1}} < 0 \, .
\end{equation}

Next, \citet{Mirouh2012} (Paper I) produced a semi-analytical model for $\gamma_{tot}^{-1}$ by developing an empirically motivated prescription for ${\rm Nu}_T-1$ given by,
\begin{equation}
{\rm Nu}_T-1 = \left( 0.75 \pm 0.05 \right) \left( \frac{{\rm Pr}}{\tau} \right)^{0.25 \pm 0.15} \frac{1-\tau}{R_0^{-1} - 1}\left( 1-r \right) \, , \nonumber \\
\end{equation}
where $r$ is the quantity defined in Equation (\ref{eq:reducedR}). They also derived an expression for $\gamma_{\rm turb}^{-1}$ that depends only on parameters that can be calculated through a linear analysis of the original governing equations in (\ref{eq:GovEqns}), 
\begin{equation}
\gamma_{\rm turb}^{-1} = \frac{\langle \tilde{w} \tilde{\mu} \rangle}{\langle \tilde{w} \tilde{T} \rangle} = R_0^{-1}\frac{\left( \lambda_{\rm R}+l^2 \right)+\lambda_{\rm I}^2}{\left( \lambda_{\rm R}+\tau l^2 \right)+\lambda_{\rm I}^2} \frac{\lambda_{\rm R} + \tau l^2}{\lambda_{\rm R} + l^2} \, ,
\end{equation}
where $\lambda_{\rm R}$ and $\lambda_{\rm I}$ are the real and imaginary parts of the growth rate of the fastest growing mode of the primary ODDC instability and $l$ is the horizontal wavenumber of the fastest growing mode (see their appendix for an explanation on how $\lambda_{\rm R}$ and $\lambda_{\rm I}$ are calculated, as well as analytical approximations in the limit of small ${\rm Pr}$ and $\tau$).

Importantly, using their model for $\gamma_{\rm tot}^{-1}$, \citet{Mirouh2012} were able to make predictions about the range of $R_0^{-1}$ where layer formation is possible. The function $\gamma_{\rm tot}^{-1}$ is concave-up with a single minimum, so by identifying the value of $R_0^{-1}$ at which $\gamma_{\rm tot}^{-1}$ reaches its minimum (referred to as $R_{\rm L}^{-1}$) it is possible to identify the region of parameter space where layers form ($R_0^{-1} < R_{\rm L}^{-1}$) and where they do not ($R_0^{-1} > R_{\rm L}^{-1}$). They found that, typically 
\begin{equation}
r_{\rm L} = \frac{R_{\rm L}^{-1} - 1}{R_{\rm c}^{-1} - 1} \sim {\rm Pr}^\frac{1}{2} \, .
\end{equation}

\citet{Wood2013} (Paper II) then presented prescriptions for quantifying the thermal and compositional fluxes through layered systems. They found that the thermal and compositional Nusselt numbers can be modeled as,
\begin{eqnarray}
{\rm Nu}_T - 1 &= &{0.1}{\rm Pr}^{\frac{1}{3}}{\rm Ra}_T^{\frac{1}{3}} \quad , \nonumber \\
{\rm Nu}_{\mu} - 1 &= &{0.03}\tau^{-1}{\rm Pr}^{\frac{1}{4}}{\rm Ra}_T^{0.37} \: .
\end{eqnarray}
The parameter ${\rm Ra}_T$ is the thermal Rayleigh number for layered convection, and is defined as a function of the layer height, $H$,
\begin{equation}
{\rm Ra}_T(H) = \left( \frac{H}{d} \right)^4 = \frac{\alpha g \left| T_{0z} - T_{0z}^{\rm ad} \right| H^4}{\kappa_T \nu} \, .
\end{equation}
It is not clear \textit{a priori} what the value of $H$ should be because in simulations of layered systems, layers always gradually merge until only a single interface remains. This suggests that some other physical mechanism outside the scope of our model of ODDC determines layer height.  For now, it is left as a free parameter, much like the mixing length in mixing-length theory \citep[see also][]{Moore2015}

By combining the results from \citet{Mirouh2012} and \citet{Wood2013} with the work done in this paper, we therefore define the effective diffusivities of temperature and chemical composition for the entire ODDC parameter regime,
\begin{eqnarray}
D_{{\rm eff},\mu} &= &\left\{ \begin{array}{ccc} \left( {0.03}\tau^{-1}{\rm Pr}^{\frac{1}{4}}{\rm Ra}_T^{0.37} + 1 \right)\kappa_{\mu} & , & R_0^{-1} < R_{\rm L}^{-1} \\ \kappa_{\mu} & , & R_0^{-1} > R_{\rm L}^{-1} \end{array} \right. \, , \nonumber \\
D_{{\rm eff},T} &= &\left\{ \begin{array}{ccc} \left( {0.1}{\rm Pr}^{\frac{1}{3}}{\rm Ra}_T^{\frac{1}{3}} + 1 \right)\kappa_T & , & R_0^{-1} < R_{\rm L}^{-1} \\ \kappa_T & , & R_0^{-1} > R_{\rm L}^{-1} \end{array} \right. \, .
\end{eqnarray}

The implication of our findings from this paper for planetary modeling is that non-layered ODDC leads to fluxes that are not significantly larger than thermal conduction or molecular diffusion. The thermal and compositional transport due to gravity waves is particularly small when compared to the increase in fluxes caused by the emergence of thermo-compositional layers (where turbulent fluxes can be orders of magnitude larger the diffusive transport). Consequently, it is not sufficient simply to know if regions in the interior of a giant planet are unstable to ODDC. The fact that layered and non-layered ODDC lead to very different transport characteristics means that special attention must be paid to calculating the threshold $R_{\rm L}^{-1}$, to determine which type of behavior will manifest.

The dynamics of gravity-wave-dominated ODDC is not expected to be pertinent to stars where typical values of $R_0^{-1}$ are in the layering regime. However, it may be important in giant planets, where higher values of {\rm Pr} (compared to stars) indicate a smaller range of $R_0^{-1}$ that is ODDC unstable, and a lower layering threshold, $R_{\rm L}^{-1}$.

In particular, the near-zero luminosity of Uranus suggests that thermal transport through the planet's interior is inefficient \citep{Hubbard1995}. Advances in equation of state research \citep{Redmer2011} lends credence to the idea that convection is being inhibited by steep compositional gradients. If this is the case, the inefficient, gravity wave-dominated, ODDC discussed in this paper may potentially play a role in Uranus's thermal evolution.

Also, though it is not yet known whether this phenomenon could produce observable signatures, the intermittent growth of shear layers discussed in this work is a potentially significant feature of non-layered ODDC in the deep interiors of giant planets. In contrast to our simulations where there is symmetry between the $x$ and $y$ directions, global rotation may provide a preferred direction for shearing motions, which could induce large scale azimuthal flows.

Future studies of ODDC will involve developing more sophisticated models through the inclusion of additional physical processes. The natural next step is to include global rotation to analyze the effects of Coriolis forces on layer formation and transport properties. This may be of particular interest because the gas giant planets in our own solar system are rapid rotators, suggesting that giant planets outside our solar system may be as well.

\acknowledgments
The authors thank J. Brown, N. Brummell, G. Glatzmaier and J. Fortney for enlightening discussions. P. G. and R. M. are funded by NST-AST-0807672 and NSF-AST-1211394. The simulations were run on the Hyades cluster, purchased using an NSF MRI grant.

\bibliographystyle{apj}

\begin{thebibliography}{18}
\expandafter\ifx\csname natexlab\endcsname\relax\def\natexlab#1{#1}\fi

\bibitem[{Baines \& Gill(1969)}]{baines1969}
Baines, P., \& Gill, A. 1969, J. Fluid Mech., 37

\bibitem[{{Hubbard} {et~al.}(1995)}]{Hubbard1995}
{{Hubbard}, W.~B. and {Podolak},M. and {Stevenson},D.~J.} 1995, {Neptune and
  Triton}, ed. {{Kruikshank}, D.~P.} (Univ. of Arizona Press, Tucson), 109

\bibitem[{{Kato}(1966)}]{kato1966}
{Kato}, S. 1966, PASJ, 18, 374

\bibitem[{{Langer} {et~al.}(1985){Langer}, {El Eid}, \& {Fricke}}]{langer1985}
{Langer}, N., {El Eid}, M.~F., \& {Fricke}, K.~J. 1985, \aap, 145, 179

\bibitem[{{Merryfield}(1995)}]{merryfield1995}
{Merryfield}, W.~J. 1995, {ApJ}, 444, 318

\bibitem[{{Mirouh} {et~al.}(2012){Mirouh}, {Garaud}, {Stellmach}, {Traxler}, \&
  {Wood}}]{Mirouh2012}
{Mirouh}, G.~M., {Garaud}, P., {Stellmach}, S., {Traxler}, A.~L., \& {Wood},
  T.~S. 2012, ApJ, 750, 61

\bibitem[{{Moore} \& {Garaud}(2015)}]{Moore2015}
{Moore}, K., \& {Garaud}, P. 2015, ArXiv e-prints

\bibitem[{Radko(2003)}]{radko2003mlf}
Radko, T. 2003, J. Fluid Mech., 497, 365

\bibitem[{{Redmer} {et~al.}(2011){Redmer}, {Mattsson}, {Nettelmann}, \&
  {French}}]{Redmer2011}
{Redmer}, R., {Mattsson}, T.~R., {Nettelmann}, N., \& {French}, M. 2011,
  \icarus, 211, 798

\bibitem[{{Rosenblum} {et~al.}(2011){Rosenblum}, {Garaud}, {Traxler}, \&
  {Stellmach}}]{rosenblum2011}
{Rosenblum}, E., {Garaud}, P., {Traxler}, A., \& {Stellmach}, S. 2011, {ApJ},
  731, 66

\bibitem[{{Schmalzl} {et~al.}(2004){Schmalzl}, {Breuer}, \&
  {Hansen}}]{schmalzl2004}
{Schmalzl}, J., {Breuer}, M., \& {Hansen}, U. 2004, EPL (Europhysics Letters),
  67, 390

\bibitem[{{Schwarzschild} \& {H{\"a}rm}(1958)}]{schwartzchildharm1958}
{Schwarzschild}, M., \& {H{\"a}rm}, R. 1958, {ApJ}, 128, 348

\bibitem[{{Spiegel} \& {Veronis}(1960)}]{spiegelveronis1960}
{Spiegel}, E.~A., \& {Veronis}, G. 1960, ApJ, 131, 442

\bibitem[{{Stevenson}(1982)}]{stevenson1982}
{Stevenson}, D.~J. 1982, \planss, 30, 755

\bibitem[{{Traxler} {et~al.}(2011){Traxler}, {Stellmach}, {Garaud}, {Radko}, \&
  {Brummell}}]{traxler2011}
{Traxler}, A., {Stellmach}, S., {Garaud}, P., {Radko}, T., \& {Brummell}, N.
  2011, ArXiv e-prints

\bibitem[{{van der Poel} {et~al.}(2013){van der Poel}, {Stevens}, \&
  {Lohse}}]{vanderpoel2013}
{van der Poel}, E.~P., {Stevens}, R.~J.~A.~M., \& {Lohse}, D. 2013, Journal of
  Fluid Mechanics, 736, 177

\bibitem[{Walin(1964)}]{walin1964}
Walin, G. 1964, Tellus, 16, 389

\bibitem[{{Wood} {et~al.}(2013){Wood}, {Garaud}, \& {Stellmach}}]{Wood2013}
{Wood}, T.~S., {Garaud}, P., \& {Stellmach}, S. 2013, ApJ, 768, 157

\end{thebibliography}

\end{document}